\documentstyle[twocolumn,aps,epsf]{revtex}
\begin{document}
\draft
\twocolumn[\hsize\textwidth\columnwidth\hsize\csname
@twocolumnfalse\endcsname


\title{Superconductivity and Antiferromagnetism \\
in Three-Dimensional Hubbard model}

\author{Tetsuya Takimoto}
\address{Advanced Science Research Center, Japan Atomic Energy 
Research Institute, \\Tokai, Ibaraki 319-1195, Japan}
\author{T$\rm{\hat o}$ru Moriya}
\address{Department of Physics, Faculty of Science and Technology, 
Science University of Tokyo, \\Noda  278-8510, Japan}

\date{\today}
\maketitle


\begin{abstract}
Interplay between antiferromagnetism and superconductivity is 
studied by using the 3-dimensional nearly half-filled 
Hubbard model with anisotropic transfer matrices $t_{\rm z}$ 
and $t_{\perp}$. 
The phase diagrams are calculated for varying values of 
the ratio $r_{\rm z}=t_{\rm z}/t_{\perp}$ using the spin 
fluctuation theory within the fluctuation-exchange approximation. 
The antiferromagnetic phase around the half-filled electron density 
expands while the neighboring phase of the anisotropic 
$d_{x^{2}-y^{2}}$-wave superconductivity shrinks with increasing 
$r_{\rm z}$. 
For small $r_{\rm z}$ $T_{\rm c}$ decreases slowly with increasing 
$r_{\rm z}$. 
For moderate values of $r_{\rm z}$ we find the second order 
transition, with lowering temperature, from the $d_{x^{2}-y^{2}}$-wave 
superconducting phase to a phase where incommensurate SDW 
coexists with $d_{x^{2}-y^{2}}$-wave superconductivity. 
Resonance peaks as were discussed previously for 2D superconductors 
are shown to survive in the $d_{x^{2}-y^{2}}$-wave superconducting 
phase of 3D systems. 
Soft components of the incommensurate SDW spin fluctuation mode 
grow as the coexistent phase is approached. 
\end{abstract}


\vskip2pc]
\narrowtext

\section{Introduction}
Unconventional superconductivity in strongly correlated 
electron systems has been one of the central issues 
in the field of condensed matter physics. 
It is widely accepted that superconductivity takes place around 
the antiferromagnetic phase for many strongly correlated systems 
such as high temperature superconductors \cite{highTc}, 
$\kappa$-BEDT organic compounds \cite{Jerome,Ishiguro}, 
and some of the heavy fermion systems \cite{Steglich,Geibel}. 
Naively, superconductivity in these systems is induced 
by the antiferromagnetic spin fluctuation \cite{Mathur}. 
In fact, the spin fluctuation theory successfully 
explains $d_{x^{2}-y^{2}}$-wave superconductivity 
observed in various quasi two-dimensional systems 
\cite{Scalapino,Moriya}. 
Recently, attractive experimental results were presented concerning 
the relation between antiferromagnetism and superconductivity. 

For example, ${\rm CeIn}_3$ with a cubic crystal structure exhibits 
an antiferromagnetic transition with decreasing temperature and 
it undergoes a superconducting transition 
under sufficient pressure 25 kbar 
where the superconducting transition temperature is 
$T_{\rm c}\approx{0.25}$ K \cite{CeIn3}. 
In relation to this system, a new heavy fermion compound 
${\rm CeRhIn}_5$ has been discovered recently, 
in which alternating layers ${\rm CeIn}_3$ and ${\rm RhIn}_2$ 
are stacked along the $c$-axis and superconductivity is 
induced by application of hydrostatic pressure \cite{CeRhIn5}. 
It is worthwhile to note that
superconducting transition temperature 
$T_{\rm c}\approx{2.1}$ K of this compound is considerably higher 
than that of ${\rm CeIn}_3$ \cite{CeRhIn5}. 
Adding to this fact, 
${\rm CeIrIn}_5$ and ${\rm CeCoIn}_5$ with the same crystal structure 
as ${\rm CeRhIn}_5$ exhibit superconductivity 
at ambient pressure \cite{CeIrIn5,CeCoIn5}, 
so that the crystal structure of ${\rm CeRhIn}_5$ is considered 
to be more favorable 
for superconductivity than the cubic one. 
These facts seem to indicate the importance of dimensionality 
for the occurence of the unconventional superconductivity. 

In another recent experiment, NQR measurement under pressure 
on CeCu$_{2}$Si$_{2}$ indicated an antiferromagnetic instability 
in the superconducting phase \cite{Ishida,Kawasaki} or possible 
coexistence between superconductivity and antiferromagnetism. 
It should also be mentioned that the observation of a neutron 
resonance peak such as those observed in high-$T_{\rm c}$ cuprates 
was reported for a heavy electron system UPd$_{2}$Al$_{3}$ 
\cite{Metoki1,Metoki2,Bernhoeft1,Bernhoeft2,Sato}. 
Thus it is worth while to study the interplay between 
antiferromagnetism and superconductivity with varying degree of 
crystal anisotropy or dimensionality 
using the spin fluctuation theory. 

The effect of dimensionality on spin fluctuation-mediated 
superconductivity was studied previously both with the use of 
phenomenological models for spin fluctuations 
\cite{Nakamura,Monthoux1} 
and from fully microscopic calculations based on the Hubbard model 
\cite{Arita}. 
In these studies comparisons were made between 
two ideal cases with simple square and cubic lattices. 
In view of recent experiments it seems important to study the effect 
of crystal anisotropy interpolating 
between ideal 2D and 3D systems. 

In this paper, we discuss the phase diagram of the three dimensional 
anisotropic Hubbard model using the fluctuation exchange (FLEX) 
approximation \cite{Bickers,Pao,Monthoux2,Dahm1}. 
The crystal anisotropy or dimensionality is controlled 
by a parameter $r_{z}$ which is the ratio of the out-of-plane 
hopping integral to the in-plane one. 
Notice that cases of 
$r_{z}=0$ and $r_{z}=1$ correspond to the square lattice 
and the cubic lattice, respectively. 
We restrict ourselves near the half-filled electron density 
and $0<r_{z}<1$. 
Our results are summarized in following four features. 
(1) With increasing $r_{\rm z}$ the area of the magnetic phase 
in the phase diagram tends to expand while 
that of the $d_{x^{2}-y^{2}}$-wave superconducting phase shrinks. 
(2) For a moderate value of $r_{\rm z}$ the $d_{x^{2}-y^{2}}$-wave 
superconducting phase 
undergoes a second order phase transition into a coexistent phase 
between $d_{x^{2}-y^{2}}$-wave superconductivity 
and incommensurate spin density wave (ICSDW). 
(3) For the same value of $r_{\rm z}$ the spin fluctuation frequency 
spectra show a resonance peak or ridge around the antiferromagnetic 
wave vector, similarly to the one studied previously for 
2-dimensional systems. 
The latter has been interpreted as the spin-excitonic collective mode 
and was assigned to the neutron resonance peaks observed in Y- and 
Bi-based high-$T_{\rm c}$ cuprates. 
(4) Low frequency components of the spin fluctuation spectrum 
corresponding to the ICSDW ordering vector grow significantly as 
the phase transition point is approached. 

In what follows we discuss the model Hamiltonian and approximation 
procedure in Sec. II and the results of calculation in Sec. III. 
Sec. IV is devoted for general discussion and summary. 


\section{Model and FLEX approximation}
\subsection{Model Hamiltonian}
The model Hamiltonian we use here is;
\begin{equation}
  H=\sum_{{\bf k},\sigma}\epsilon_{{\bf k}}
  a_{{\bf k}\sigma}^{\dagger}a_{{\bf k}\sigma}
  +U\sum_{i}n_{i\uparrow}n_{i\downarrow} ,
\end{equation}
where $a_{{\bf k}\sigma}^{\dagger}$ is the creation operator for 
a quasi-particle with momentum ${\bf k}$ and spin $\sigma$, $U$ is 
the on-site Coulomb energy, $n_{i\sigma}$ is the number operator for 
a quasi-particle with spin $\sigma$ at $i$-site. 
$\epsilon_{{\bf k}}$ is the energy 
dispersion of the quasi-particle given by
\begin{eqnarray}
    \epsilon_{{\bf k}}=&&
     -2t(\cos{k_{x}}+\cos{k_{y}})+4t'\cos{k_{x}}\cos{k_{y}}\nonumber\\
    &&-2t\hspace{1mm}r_{z}\cos{k_{z}}
      +4t'r_{z}\cos{k_{z}}(\cos{k_{x}}+\cos{k_{y}})
\end{eqnarray}
where $t$ and $t'$ are the nearest-neighbor and 
the next nearest-neighbor hopping integrals, respectively. 
Hereafter, $t$ is used as the unit of energy. 
Also, we introduce a parameter $r_{z}$ describing the anisotropy 
along $z$-axis where $r_{z}=0$ corresponds to the square lattice, 
and $r_{z}=1$ to the cubic lattice. 
Although the value of the anisotropy parameter 
for the next nearest-neighbor hopping should 
be generally different from that for the nearest-neighbor, 
we use the same value for them since no essential difference 
will appear by definite distinction between them. 
The variation of the phase diagram with this anisotropy is studied 
using the Green's function method. 

\subsection{FLEX Approximation}
When the interaction constant $U$ is equal to zero, 
the one-particle Green's function 
of the quasi-particle at temperature $T$ is given by 
\begin{equation}
  G^{(0)}({\bf k},{\rm i}\omega_{n})=
  \frac{1}{{\rm i}\omega_{n}-\epsilon_{\bf k}+\mu}
\end{equation}
where $\omega_{n}=(2n+1)\pi T$ is the Fermion Matsubara frequency 
and $\mu$ is the chemical potential. 

For the three dimensional system with $U\neq0$, 
various ordered states appear at respective transition temperature 
below which features of the system is described by 
corresponding order parameters, and anomalous Green's functions. 
In order to study the interplay between magnetism and 
superconductivity, we wish to construct the phase diagram 
of this model calculating the magnetic and superconducting transition 
points. 
In the FLEX approximation the superconducting $T_{\rm c}$ is 
calculated within a mean field level while the magnetic susceptibility 
is renormalized self-consistently and thus there is 
no Neel temperature $T_{\rm N}$ for 2D-systems. 
Here the finite value of $r_{\rm z}$ or 3D-character makes it 
possible to find magnetic transition point $T_{\rm N}$ from 
the divergence of $\chi_{Q}$. 

The Dyson-Gor'kov equations for 
the Green's functions and the anomalous Green's functions 
are given by 
\begin{eqnarray}
  &&G(k)=G^{(0)}(k)+G^{(0)}(k)\Sigma^{(1)}(k)G(k)\nonumber\\
  &&\hspace*{24mm}-G^{(0)}(k)\Sigma^{(2)}(k)F^{\dagger}(k)\\
  &&F^{\dagger}(k)=G^{(0)}(-k)\Sigma^{(1)}(-k)F^{\dagger}(k)\nonumber\\
  &&\hspace*{10mm}+G^{(0)}(-k)\Sigma^{(2)}(-k)G(k)
\end{eqnarray}
where the abbreviation $k\equiv({\bf k},{\rm i}\omega_{n})$ is used 
and $G(k)$ is Green's function for $U\neq0$. 
$F^{\dagger}(k)$ is anomalous Green's function to describe 
the superconducting phase. 
The self-energies $\Sigma^{(1)}(k)$ and $\Sigma^{(2)}(k)$ are given 
within the FLEX approximation as follows 
\cite{Bickers,Pao,Monthoux2,Dahm1}:
\begin{eqnarray}
  &&\Sigma^{(1)}(k)=\sum_{q}V_{\rm eff}(q)G(k-q),\\
  &&\Sigma^{(2)}(k)=-\sum_{q}V_{\rm sing}(q)F^{\dagger}(k-q),\\
  &&V_{\rm eff}(q)=U^{2}[\frac{3}{2}\chi^{\rm s}(q)
                        +\frac{1}{2}\chi^{\rm c}(q)
                        -\frac{1}{2}\{\overline{\chi}^{\rm s}(q)
                                     +\overline{\chi}^{\rm c}(q)\}]\\
  &&V_{\rm sing}(q)=U^{2}[\frac{3}{2}\chi^{\rm s}(q)
                         -\frac{1}{2}\chi^{\rm c}(q)
                         -\frac{1}{2}\{\overline{\chi}^{\rm s}(q)
                                      -\overline{\chi}^{\rm c}(q)\}]
\end{eqnarray}
with
\begin{eqnarray}
  &&\chi^{\rm s}(q)=\frac{\overline{\chi}^{\rm s}(q)}
                         {1-U\overline{\chi}^{\rm s}(q)},\hspace{5mm}
    \chi^{\rm c}(q)=\frac{\overline{\chi}^{\rm c}(q)}
                         {1+U\overline{\chi}^{\rm c}(q)},\\
  &&\overline{\chi}^{\rm s(c)}(q)=-\sum_{k}
    [G(k+q)G(k){\pm}F^{\dagger}(k+q)F^{\dagger}(k)]
\end{eqnarray}
where $q\equiv({\bf q},{\rm i}\Omega_{n})$ 
and $\Omega_{n}=2n\pi T$ is the Boson Matsubara frequency. 
Also, $\sum_{k}\equiv\frac{T}{N_{0}}\sum_{\bf k}\sum_{n}$ 
and $N_{0}$ is the number of sites. 
$\chi^{\rm s(c)}(q)$ and $\overline{\chi}^{\rm s(c)}(q)$ are 
spin (charge) susceptibility and its irreducible part, respectively. 
Due to the analytical continuation of $\chi^{\rm s(c)}(q)$ 
from the imaginary axis to the real axis, spectrum of spin (charge) 
fluctuations is provided. 

For the purpose of calculating $T_{\rm c}$ we linearize 
this Dyson-Gor'kov equations with respect to $F^{\dagger}(k)$ 
or $\Sigma^{(2)}(k)$ as follows:
\begin{eqnarray}
  &&G(k)=\frac{1}{G^{(0)}(k)^{-1}
               -\Sigma^{(1)}(k)},\\
  &&F^{\dagger}(k)=|G(k)|^{2}\Sigma^{(2)}(-k)
\end{eqnarray}
where the normal self-energy $\Sigma^{(1)}(k)$ in eq. (12) 
corresponds to that of the normal state. 
The transition temperature for superconductivity is determined 
as the temperature below which the linearized equation 
for $\Sigma^{(2)}(k)$ has a non-trivial solution. 
The linearized equation is given as
\begin{equation}
  \Sigma^{(2)}(k)=-\sum_{p}[V_{\rm sing}(k-p)
    |G(p)|^{2}]_{\Sigma^{(2)}\rightarrow 0}\Sigma^{(2)}(p)
\end{equation}
where $T_{\rm c}$ is obtained by the temperature at which the maximum 
eigenvalue becomes unity. 
Note that the superconducting order parameter has 
the same symmetry as $\Sigma^{(2)}(k)$. 

\subsection{Details of Numerical Calculation}
The FLEX calculation is numerically carried out for each value 
of $r_{z}$ at fixed parameter values of $t'=0.2$ and $U=6$. 
All summations involved in the above self-consistent equations 
are calculated using FFT algorism for the ${\bf k}$-space 
with $16\times16\times16$ meshes in the first Brillouin zone 
and Matsubara frequency 
sums with sufficiently large cutoff-frequency $\omega_{\rm c}=36$. 
Although the lattice size seems to be 
relatively small, it is considered that the obtained results 
are quantitatively modified but does not qualitatively change 
with magnification of the system size. 
When relative error of the self-energy for all ${\bf k}$ and 
$\omega_{n}$ becomes smaller than $10^{-6}$, it is assumed that 
the solution is obtained for the self-consistent equations 
mentioned above. 
The second-order magnetic phase transition is given by 
\begin{equation}
  U\overline{\chi}^{\rm s}({\bf q},{\rm i}\Omega_{n}=0)=1
\end{equation}
where a spin susceptibility 
${\chi}^{\rm s}({\bf q},{\rm i}\Omega_{n}=0)$ diverges. 
Since numerical calculation can not treat any divergences, 
we always adopt following condition as the condition for the 
magnetic transition 
\begin{equation}
  U\overline{\chi}^{\rm s}({\bf q},{\rm i}\Omega_{n}=0)=1-\epsilon
\end{equation}
where $\epsilon=0.002$ is used throughout these calculations; 
$U\overline{\chi}^{\rm s}({\bf q},0)\approx1$\cite{footnote}. 
It should be noted that slight modification for the value 
of $\epsilon>0$ 
does not bring about qualitative changes for our results. 
The superconducting transition tempertures are determined by 
solving the gap equation (14) obtained within FLEX approximation. 
The analytical continuations of $\chi^{\rm s(c)}(q)$ to the real axis 
are carried out by using Pad$\acute{\rm e}$ approximants.


\section{Calculated Results}
\subsection{Phase Diagrams}
In this section, the results of FLEX calculation 
for the model mentioned above are shown. 
We begin by examining variation with the anisotropy 
parameter $r_{z}$ of the phase diagram in temperature $T$ vs 
hole-doping concentration $(1-n)$ space, 
where $n$ is the electron density per site. 
The calculated phase diagrams are shown in the Fig. 1; 
1(a), 1(b), 1(c), 1(d), and 1(e) are the results 
for $r_{z}=0.1$, 0.2, 0.4, 0.6, and 0.8, respectively. 
Each symbol used in these figures has the following meaning; 
closed circles show the transition to the superconducting phase 
with $B_{\rm 1g}$-symmetry and 
open circles, open squares, open diamonds, and 
open triangles correspond to the magnetic transition 
with the SDW wave vector of ${\bf Q}=(\pi,\pi,\pi)$, 
${\bf Q}_{1}=(0.88\pi,\pi,\pi)$, ${\bf Q}_{2}=(0.75\pi,\pi,\pi)$, 
and ${\bf Q}_{3}=(0.88\pi,0.88\pi,\pi)$, respectively. 
Also, closed squares in Fig. 1(b) and closed diamonds 
in Fig. 1(c) show {\it the second-order magnetic transition 
between the $d_{x^{2}-y^{2}}$-wave superconducting phase 
and the coexistent phase} 
with the SDW wave vector 
of ${\bf Q}_{1}$ and ${\bf Q}_{2}$, respectively. 
It should be noted that all of these phase transitions are 
of the second-order. 
Adding to these phase transition points, 
symbols of plus, times, sharp, and asterisk in Fig. 1 
correspond to the crossover temperatures at which 
the peak position of 
$\chi^{\rm s}({\bf q},0)$ changes from ${\bf Q}$ to ${\bf Q}_{1}$, 
${\bf Q}_{1}$ to ${\bf Q}_{2}$, ${\bf Q}_{1}$ to ${\bf Q}_{3}$, 
and ${\bf Q}_{3}$ to ${\bf Q}_{2}$, respectively. 
For example, in the case of $n=0.85$ in Fig. 1(b), 
the peak position of $\chi^{\rm s}({\bf q},0)$ changes from 
${\bf Q}$ to ${\bf Q}_{1}$ with the decrease of temperature, 
and then $d_{x^{2}-y^{2}}$-wave superconductivity appears. 
For more detailed change in the peak position we need more 
detailed calculation using smaller mesh in the ${\bf q}$-space. 
In the systems with nearly half-filled density, the peak of 
$\chi^{\rm s}({\bf q},0)$ always appear 
at the commensurate position ${\bf q=Q}$ 
for moderately high temperature region as shown in Fig. 1. 
This fact is consistent with the results of previous 
calculations for 2D systems \cite{Pao,Monthoux2,Dahm1}.

\begin{figure}[t]
\centerline{\epsfxsize=8.5truecm \epsfbox{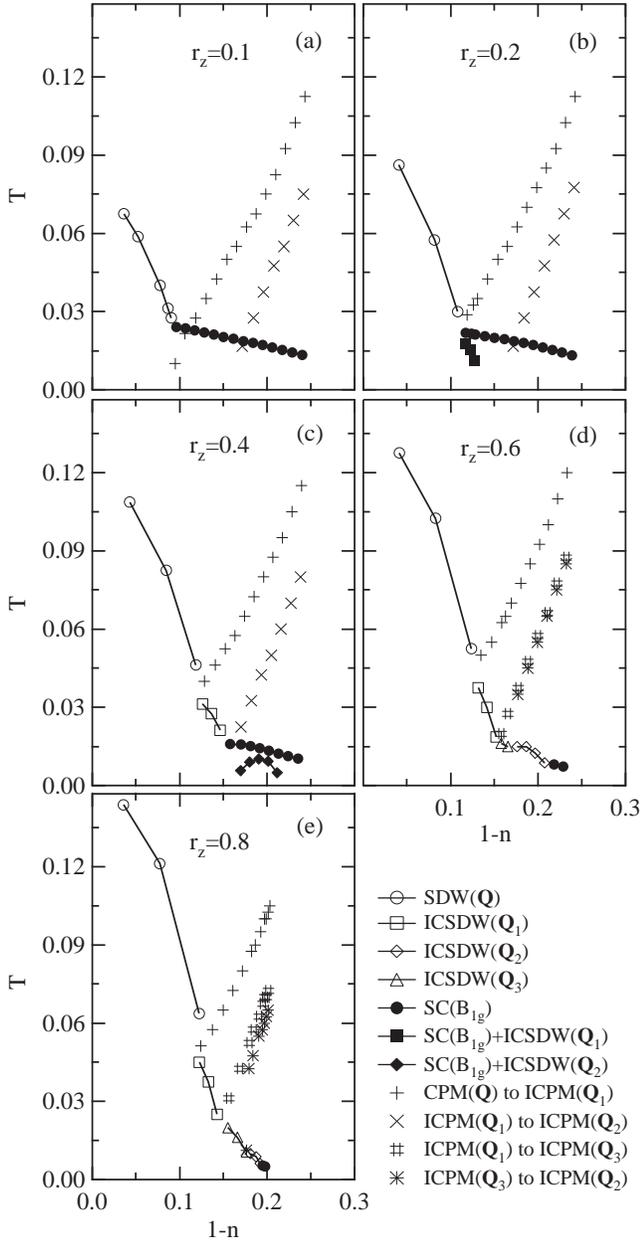} }
\label{fig1}
  \caption
  {$T$ vs $1-n$ phase diagrams for (a) $r_{z}=0.1$, 
  (b) $r_{z}=0.2$, (c) $r_{z}=0.4$, (d) $r_{z}=0.6$, and 
  (e) $r_{z}=0.8$. The abbreviations used here have following 
  meaning; SDW(${\bf Q}$): commensurate SDW phase, 
  ICSDW(${\bf Q}_{n}$): incommensurate SDW phase 
  with wave vector ${\bf Q}_{n}$, 
  SC($B_{\rm 1g}$): $d_{x^{2}-y^{2}}$-wave superconducting phase, 
  CPM(${\bf Q}$): paramagnetic phase with commensurate peak for 
  $\chi^{\rm s}({\bf q},0)$, and 
  ICPM(${\bf Q}_{n}$): paramagnetic phase with incommensurate peak 
  at ${\bf q}={\bf Q}_{n}$. Also, the meanings of all symbols 
  is explained in text.}
\end{figure}

One of the important results we find from Fig. 1 is the gradual 
suppression of the $d_{x^{2}-y^{2}}$-wave superconducting phase 
with increasing $r_{\rm z}$ 
or increasing three dimensionality in contrast to 
the gradual expansion of the antiferromagnetic phase. 
For $r_{\rm z}=0.8$ the $d_{x^{2}-y^{2}}$-wave superconducting phase 
is very much suppressed 
in accord with the previous calculation for the simple cubic lattice 
\cite{Arita}. 
An interesting result here is that with increasing $r_{\rm z}$ 
from zero or 2D limit $T_{\rm c}$ first decreases but slowly. 
Even for $r_{\rm z}=0.4$, with substantial 3D character, $T_{\rm c}$ 
is not much reduced from the 2D value. 
Explicit plots are found in the inset of Fig. 2. 
For still increasing value of $r_{\rm z}$, $T_{\rm c}$ decreases 
rapidly. 
These results indicate that materials with layered structure 
such as cuprates and Ce-115 heavy fermion compounds are quite 
favorable for the appearance of $d_{x^{2}-y^{2}}$-wave 
superconductivity.

\begin{figure}[t]
\centerline{\epsfxsize=8.5truecm \epsfbox{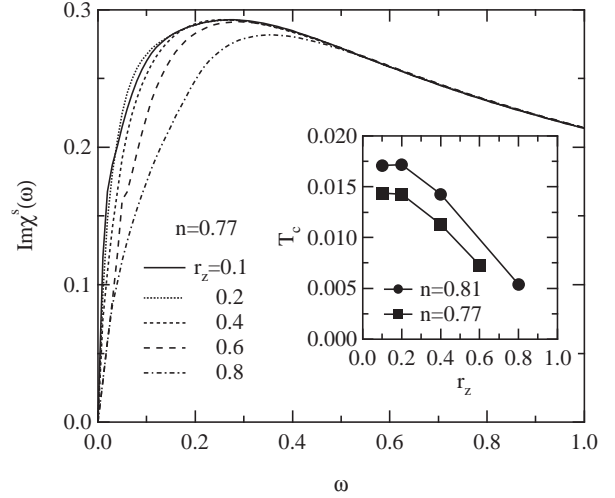} }
\label{fig2}
  \caption
  {Calculated results of $\omega$-dependences of 
  ${\rm Im}\chi^{\rm s}(\omega)$ 
  for respective $r_{z}$-value at fixed electron density $n=0.77$ 
  where temperature is set just above $T_{\rm c}$ except for 
  the case of $r_{z}=0.8$ for which $T=0.005$. 
  Inset: $r_{z}$-dependence of $T_{\rm c}$ 
  at fixed electron densities $n=0.81$ and $n=0.77$. }
\end{figure}

Another important result is the possible coexistent phase between 
$d_{x^{2}-y^{2}}$-wave superconductivity and ICSDW 
as seen in Fig. 1(b) and 1(c). 
It should 
be noted that the SDW ordering vectors ${\bf Q}_{1}$ and ${\bf Q}_{2}$ 
in 1(b) and 1(c) are the same as the ones corresponding to the peak 
positions of $\chi^{\rm s}({\bf q},0)$ just above the superconducting 
transition temperature. 
Considering this new feature of the phase diagram it seems worth while 
to look at the dynamical susceptibility around the coexistent phase. 
It is also interesting to see if the spin-excitonic collective modes 
as were found in 2D systems persist for 3D systems 
and give simultaneous explanations 
for the resonance peaks observed in high-$T_{\rm c}$ cuprates and 
in certain heavy electron superconductor. 
These problems will be discussed in the following subsection.

\subsection{Spectra of Spin Fluctuations}
In order to get some idea about the possible reason for the 
reduction of $T_{\rm c}$ with increasing $r_{\rm z}$ 
we have calculated the ${\bf q}$-integrated imaginary part of 
the dynamical susceptibility
\begin{equation}
  {\rm Im}\chi^{\rm s}(\omega)
  =\frac{1}{N_0}\sum_{\bf q}
    {\rm Im}\chi^{\rm s}({\bf q}, \omega+{\rm i}\delta)
\end{equation}
for varying values of $r_{\rm z}$ at a fixed electron density 
0.77 and just above $T_{\rm c}$. 
The results are shown in Fig. 2. 
The $r_{\rm z}$-dependence of $T_{\rm c}$ is also shown in the inset 
for $n=0.77$ and $n=0.81$. 

We see from this figure that with increasing $r_{\rm z}$ 
the relatively low frequency components (but much higher than 
the order of $T_{\rm c}$) decreases while the higher frequency side 
of the spectrum is fixed. 
This decrease of the intensity of spin fluctuation is particularly 
significant for $r_{\rm z}>0.4$ and is considered to be 
responsible for the reduction of $T_{\rm c}$.

\begin{figure}[t]
\centerline{\epsfxsize=8.5truecm \epsfbox{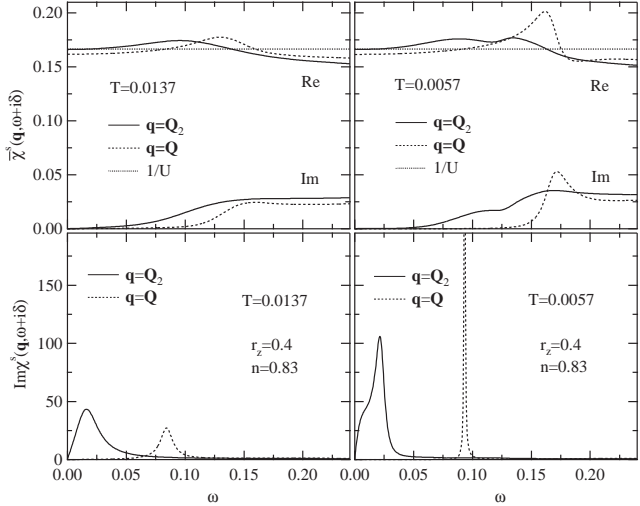} }
\label{fig3}
  \caption
  {Calculated results of $\omega$-dependences of 
  (upper) $\overline{\chi}^{\rm s}({\bf q}, \omega+{\rm i}\delta)$ 
  and (lower) Im$\chi^{\rm s}({\bf q}, \omega+{\rm i}\delta)$ 
  for ${\bf q}={\bf Q}_{2}$(solid line) and ${\bf q}={\bf Q}$
  (dotted line) at two temperatures (left) $T=0.0137$ and 
  (right) $T=0.0057$ below $T_{\rm c}=0.0157$ where 
  parameter set ($r_{z}, n$) is fixed to (0.4, 0.83). }
\end{figure}

We next discuss the behaviors of the predominant components of 
the dynamical susceptibility in the situations where the phase 
transition between the $d_{x^{2}-y^{2}}$-wave superconducting 
and the coexistent phases 
takes place as seen in Fig. 1(b) and 1(c). 
As a representative example we choose the parameter set 
$(r_{\rm z}, n)=(0.4, 0.83)$ in Fig. 1(c). 
We show in Fig. 3 the spin fluctuation spectra below $T_{\rm c}$ 
for the antiferromagnetic component with the wave vector $\bf Q$ 
and for the SDW components with the incommensurate wave vector 
${\bf Q}_{2}$. 
The upper and lower pannels show the calculated results 
of the $\omega$-dependences of the irreducible spin susceptibilities 
and the imaginary part of the spin susceptibilities, respectively, 
where these quantities have following relation 
according to equation (10)
\begin{eqnarray}
  &&{\rm Im}\chi^{\rm s}({\bf q}, \omega+{\rm i}\delta)\nonumber\\
  &&=\frac{{\rm Im}\overline{\chi}^{\rm s}
                   ({\bf q}, \omega+{\rm i}\delta)}
  {[1-U{\rm Re}\overline{\chi}^{\rm s}
               ({\bf q}, \omega+{\rm i}\delta)]^{2}
  +[U{\rm Im}\overline{\chi}^{\rm s}
             ({\bf q}, \omega+{\rm i}\delta)]^{2}}.
\end{eqnarray}
We first note that the intensity of 
Im$\chi^{\rm s}({\bf Q}, \omega+{\rm i}\delta)$ 
around $\omega=0.08$ develops strongly with decreasing temperature. 
As proposed in previous papers \cite{Morr,Dahm2,Takimoto1}, 
it is considered that 
this peak corresponds to the spin-excitonic collective mode 
within the $d_{x^{2}-y^{2}}$-wave superconducting phase 
taking into account that the excitation energy for this mode 
is less than the superconducting gap 
$2\Delta\approx0.17$ estimated from the peak-position of 
Im$\overline{\chi}^{\rm s}({\bf Q}, \omega+{\rm i}\delta)$. 
Noting that relatively large value of $r_{z}$ is used in Fig. 3, 
it is considered that such a collective mode 
is not specific to 2D systems but appears 
more generally in 3D systems with 
the $d_{x^{2}-y^{2}}$-wave superconducting phase 
induced by the antiferromagnetic spin fluctuation. 
Thus, although this feature is discussed in relation with 
the 41meV resonance peak observed in Y- and Bi-based cuprates 
\cite{Rossat-Mignod,Fong1}, 
it is expected that such a collective mode within 
the $d_{x^{2}-y^{2}}$-wave superconducting state 
should be seen in another unconventional superconductors 
such as UPd$_{2}$Al$_{3}$ 
\cite{Metoki1,Metoki2,Bernhoeft1,Bernhoeft2,Sato}. 
As is seen in Fig. 3, the spectrum of the spin fluctuation 
with the wave vector ${\bf Q}_{2}$ also 
exhibits such a collective mode around $\omega=0.02$. 
In addition to this, a new excitation appears 
with decreasing temperature as a shoulder 
of ${\rm Im}\chi^{\rm s}({\bf Q}_{2}, {\omega}+{\rm i}\delta)$
near $\omega=0$ . 
This behavior is considered to be associated with the instability 
of the ICSDW mode ${\bf Q}_{2}$ at the second order transition 
point to the coexistent phase. 
Recent higher-resolution inelastic neutron scattering experiment 
shows that the shoulder on low excitation energy is observed at 
the incommensurate position below $T_{\rm c}$ in the under-doped 
high-$T_{\rm c}$ cuprates\cite{Fong2}. 
Such an experimental data may be related with 
the behavior shown in Fig. 3.

\begin{figure}[t]
\centerline{\epsfxsize=8.5truecm \epsfbox{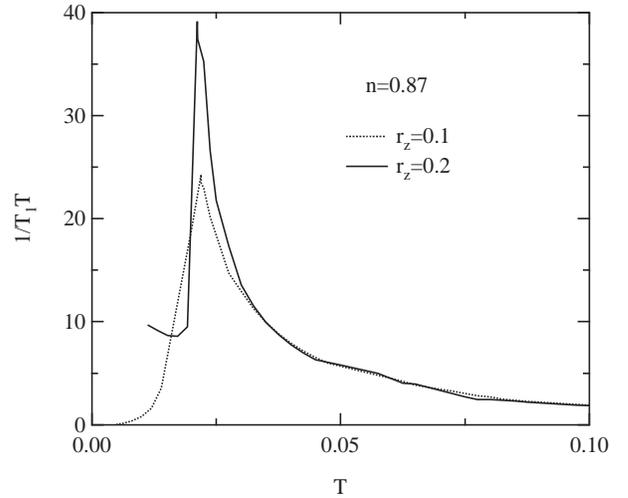} }
\label{fig4}
  \caption
  {Calculated results of $T$-dependences of $1/T_{1}T$ 
  for $r_{z}=0.1$ (dotted line) and $r_{z}=0.2$ (solid line) 
  at fixed electron density $n=0.87$ 
  where their $d_{x^{2}-y^{2}}$-wave superconducting transition 
  temperatures are almost same as seen in Fig. 1. }
\end{figure}

These low energy excitations are considered to affect 
various physical quantities, especially, 
the nuclear spin-lattice relaxation rate $1/T_{1}$ 
which is expressed as
\begin{equation}
  \frac{1}{T_{1}T}\approx\sum_{\bf q}
  \frac{{\rm Im}\chi^{\rm s}({\bf q}, \omega+{\rm i}\delta)}{\omega}
  |_{\omega\rightarrow 0}
\end{equation}
where the hyperfine coupling constant is assumed to be unity. 
It is well known that $1/T_{1}T$ exhibits the $T^2$-law 
for the $d_{x^{2}-y^{2}}$-wave superconductor due to 
the existence of line nodes as in the high-$T_{\rm c}$ cuprates. 
But, it is expected that if this type of excitations exist as seen for 
${\rm Im}\chi^{\rm s}({\bf Q}_{2}, {\omega}+{\rm i}\delta)$ in Fig. 3, 
$1/T_{1}T$ should show an upturn below $T_{\rm c}$ as shown in Fig. 4. 
In fact, such a behavior is recently reported for 
${\rm CeCu}_{2}{\rm Si}_2$ \cite{Ishida,Kawasaki} 
where it is known that the unconventional superconductiviting 
transition takes place at low temperature 
and the magnetic phase, so-called $A$-phase, exists 
under the magnetic field.


\section{Discussion and Summary}
Since the second-order phase transition accompanies 
some symmetry-breaking, it is instrucitve to discuss phase 
transitions considering the symmetry for the respective phase 
in those phase diagrams. 
As wellknown, the Hubbard model shown in eq. (1) has 
$D_{4h}{\times}SU(2){\times}\Theta{\times}U(1)$-symmetry where 
$D_{4h}$ is tetragonal point group of the system, $SU(2)$ rotational 
invariance in the spin-space, $\Theta$ time-reversal symmetry, 
and $U(1)$ gauge symmetry. 
When the antiferromagnetic or SDW transition takes place 
with the magnetic moment along $z$-axis, 
the $SU(2)$ rotational invariance is broken and 
$D_{4h}{\times}\Theta$ reduces to its subgroup $\overline{D}_{4h}$, 
so that the symmetry of the system lowers to 
$\overline{D}_{4h}{\times}U(1)$ 
where $\overline{G}$ is a magnetic point group which includes 
some group elements for $G$'s typical element $R$ to couple with 
the time-reversal operator $\theta$. 
Also, the $d_{x^{2}-y^{2}}$-wave superconducting 
transition (belonging to $B_{\rm 1g}$ irreducible representation) 
breaks $U(1)$ gauge symmetry and $\pi/2$-rotational invariance, 
so that the symmetry of this phase becomes 
$D_{2h}{\times}SU(2){\times}\Theta$. 

Of course, the first-order phase transition between SDW and 
$d_{x^{2}-y^{2}}$-wave superconducting phases is possible 
where a group of one phase is not a subgroup of the other. 
Noting that no magnetic transition takes place 
in the $d_{x^{2}-y^{2}}$-wave superconducting phase in Fig. 1(a), 
it seems that such first-order phase transitions take place 
between SDW and $d_{x^{2}-y^{2}}$-wave superconducting phases 
as schematically shown in Fig. 5(a). 
On the other hand, the second-order phase transitions are expected 
between the coexistent phase and the ICSDW and $d_{x^{2}-y^{2}}$-wave 
superconducting phases 
because the group $\overline{D}_{2h}$ of 
the coexistent phase is a subgroup of 
both of the other two phases 
as seen in Fig. 5(b). 
It seems that Fig. 1(b) corresponds to such a case. 
In Fig. 1(c), it seems that the transition between the 
$d_{x^{2}-y^{2}}$-wave superconducting phase and the coexistent 
phase is of the second-order 
while the first-order phase transition is expected between 
the $d_{x^{2}-y^{2}}$-wave superconducting and the ICSDW phases, 
as shown in Fig. 5(c). 


Finally, it may be worth while to make a comment 
on the spin excitations in the coexistent phase. 
In view of the symmetry of the coexistent phase 
we may naturally expect to have a Goldstone or 
zero frequecncy spin wave mode 
which recovers the broken $SU(2)$-symmetry. 
We have seen in Fig. 3 that the soft component as a shoulder 
of the spin excitonic collective mode at the ICSDW wave vector 
${\bf Q}_{2}$ grows while the main peak position is kept finite 
as the coexistent phase is approached. 
The softening component may be regarded as a precursor 
to the appearance of the Goldstone mode. 
Thus in the coexistent phase we may generally expect to have 
two types of spin excitation modes, 
the spin excitonic collective mode associated with the superconducting 
gap and the spin wave mode associated with antiferromagnetism 
(including SDW). 
When $T_{\rm N}$ is substantially higher than $T_{\rm c}$, however, 
the appearance of the former mode itself seems to be dubious from 
consideration of the possible gap structure. 
It should be an interesting future subject to study 
the dispersions and the ranges of appearance of these modes.

\begin{figure}[t]
\centerline{\epsfxsize=8.5truecm \epsfbox{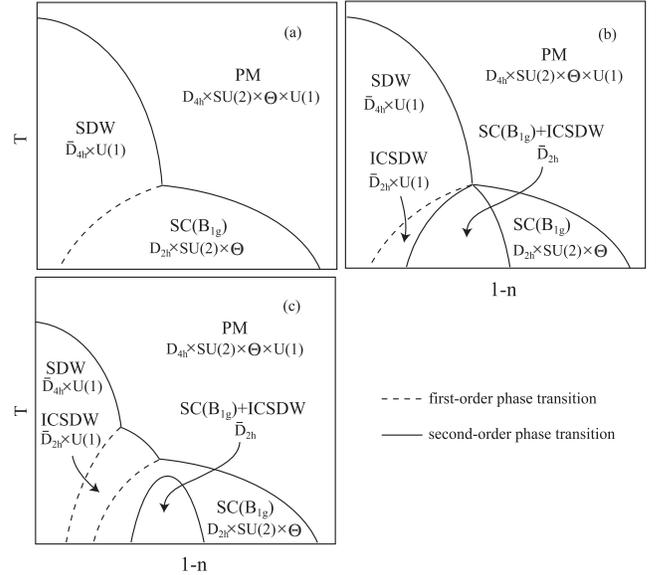} }
\label{fig5}
  \caption
  {Schematic phase diagrams corresponding to Fig. 1(a), 1(b), 
  and 1(c). The left edges of these figures correspond to 
  half filled density. Here, PM, SDW, ICSDW, and SC($B_{\rm 1g}$) 
  stand for the paramagnetic metal, SDW, incommensurate SDW, and 
  $d_{x^{2}-y^{2}}$-wave superconducting phase, respectively. 
  In SC($B_{\rm 1g}$)+ICSDW phase, $d_{x^{2}-y^{2}}$-wave 
  superconductivity and ICSDW coexist. Also, the symmetry of 
  respective phase is written in order to discuss what kind of 
  symmetry is broken. }
\end{figure}

In summary, the phase diagram of the three-dimensional anisotropic 
Hubbard model with varying ratio $r_{\rm z}$ of the transfer matrices 
along z-axis and in xy-plane is calculated by 
using the spin fluctuation theory within the FLEX approximation. 
The results are summarized as follows: 
(1) For small $r_{\rm z}$ or nearly 2-dimensional case we have 
an antiferromagnetic phase around the half-filled electron density 
and a $d_{x^{2}-y^{2}}$-wave superconducting phase neighboring to it. 
With increasing $r_{\rm z}$ the antiferromagnetic phase tends to 
expand while the $d_{x^{2}-y^{2}}$-wave superconducting phase shrinks. 
For relatively small values of $r_{\rm z}$, however, 
this trend is weak and the superconducting transition temperature 
$T_{\rm c}$ decreases slowly with increasing $r_{\rm z}$. 
As $r_{\rm z}$ approaches 1, $T_{\rm c}$ decreases rapidly. 
This behavior of $T_{\rm c}$ seems to be related with the reduction 
in the ${\bf q}$-integrated intensity of the spin fluctuation 
spectrum mainly in relatively low frequency (but much higher than 
the order of $T_{\rm c}$) side. 
(2) For a moderate value of $r_{\rm z}$ the $d_{x^{2}-y^{2}}$-wave 
superconducting phase 
undergoes a second-order transition, with decreasing temperature, 
into a phase where incommensurate spin density wave (ICSDW) 
coexists with $d_{x^{2}-y^{2}}$-wave superconductivity. 
(3) For the same value of $r_{\rm z}$ the spin fluctuation frequency 
spectra show a resonance peak or ridge around the antiferromagnetic 
wave vector, similarly to the one studied previously for 
two-dimensional systems. 
These spin-excitonic collective modes are considered to explain 
the neutron resonance peaks observed in Y- and Bi-based 
high-$T_{\rm c}$ cuprates and in a heavy Fermion system 
UPd$_{2}$Al$_{3}$. 
(4) The low frequency components of the ICSDW mode of spin 
fluctuations grow significantly as the transition point 
between the $d_{x^{2}-y^{2}}$-wave superconducting and 
coexistent phases is approached. 

Recently, in order to discuss the effect of orbital degeneracy 
to superconductivity, the weak coupling theory is developed 
where the orbital fluctuation plays an important role as well as 
spin fluctuation \cite{Takimoto2,Takimoto3}. 
Since any fluctuations should be treated dynamically, 
it is desireble to construct the strong coupling theory 
including the orbital degree of freedom.

\section*{Acknowledgements}
The authors are indebted T. Hotta, H. Kondo, S. Nakamura, 
K. Ueda, K. Yamaji and T. Yanagisawa for useful discussions. 
One of the authors (T. T.) would like to thank support from 
Japan Science and Technology Corporation, 
Domestic Research Fellow.


\begin{references}

\bibitem{highTc}
J. G. Bednorz and K. A. M$\ddot{\rm u}$ller, 
Z. Phys. B {\bf 64}, 189 (1986).

\bibitem{Jerome}
D. J\'erome and H. J. Schulz, 
Adv. Phys. {\bf 31}, 299 (1982).

\bibitem{Ishiguro}
T. Ishiguro, K. Yamaji, and G. Saito, 
{\it Organic Superconductors}, 2nd ed. (Springer-Verlag, Berlin, 1998).

\bibitem{Steglich}
F. Steglich, J. Aarts, C. D. Bredl, W. Lieke, 
D. Meschede, W. Franz, and H. Sch$\ddot{\rm a}$fer,
Phys. Rev. Lett. {\bf 43}, 1892 (1979).

\bibitem{Geibel}
C. Geibel, C. Schank, S. Thies, H. Kitazawa, 
C. D. Bredl, A. B$\ddot{\rm o}$hm, M. Rau, A. Grauel, 
R. Caspary, R. Helfrich, U. Ahlheim, G. Weber, 
and F. Steglich, 
Z. Phys. B {\bf 84}, 1 (1991).

\bibitem{Mathur}
N. D. Mathur, F. M. Grosche, S. R. Julian, 
I. R. Walker, D. M. Freye, R. K. W. Haselwimmer, 
and G. G. Lonzarich, 
Nature {\bf 394}, 39 (1998). 

\bibitem{Scalapino}
D. J. Scalapino, 
Phys. Rep. {\bf 250}, 329 (1995).

\bibitem{Moriya}
T. Moriya and K. Ueda, 
Adv. Phys. {\bf 49}, 555 (2000).

\bibitem{CeIn3}
I. R. Walker, F. M. Grosche, D. M. Freye, 
and G. G. Lonzarich, 
Physica C {\bf 282-287}, 303 (1997).

\bibitem{CeRhIn5}
H. Hegger, C. Petrovic, E. G. Moshopoulou, 
M. F. Hundley, J. L. Sarrao, Z. Fisk, 
and J. D. Thompson, 
Phys. Rev. Lett. {\bf 84}, 4986 (2000).

\bibitem{CeIrIn5}
C. Petrovic, R. Movshovich, M. Jaime, P. G. Pagliuso, 
M. F. Hundley, J. L. Sarrao, Z. Fisk, 
and J. D. Thompson, 
Europhys. Lett. {\bf 53}, 354 (2001).

\bibitem{CeCoIn5}
C. Petrovic, P. G. Pagliuso, M. F. Hundley, 
R. Movshovich, J. L. Sarrao, J. D. Thompson, 
Z. Fisk, and P. Monthoux, 
J. Phys.: Condens. Matter {\bf 13}, 
L337 (2001).

\bibitem{Ishida}
K. Ishida, Y. Kawasaki, K. Tabuchi, K. Kashima, 
Y. Kitaoka, K. Asayama, C. Geibel, and F. Steglich, 
Phys. Rev. Lett. {\bf 82}, 5353 (1999).

\bibitem{Kawasaki}
Y. Kawasaki, K. Ishida, T. Mito, C. Thessieu, 
G.-q. Zheng, Y. Kitaoka, C. Geibel, and F. Steglich, 
Phys. Rev. B{\bf 63}, 140501(R) (2001).

\bibitem{Metoki1}
N. Metoki, Y. Haga, Y. Koike, N. Aso, and Y. $\bar{\rm O}$nuki, 
J. Phys. Soc. Jpn. {\bf 66}, 2560 (1997).

\bibitem{Metoki2}
N. Metoki, Y. Haga, Y. Koike, and Y. $\bar{\rm O}$nuki, 
Phys. Rev. Lett. {\bf 80}, 5417 (1998).

\bibitem{Bernhoeft1}
N. Bernhoeft, N. Sato, B. Roessli, N. Aso, A. Hiess, 
G. H. Lander, Y. Endoh, and T. Komatsubara, 
Phys. Rev. Lett. {\bf 81}, 4244 (1998).

\bibitem{Bernhoeft2}
N. Bernhoeft, 
Eur. Phys. J. B {\bf 13}, 685 (2000).

\bibitem{Sato}
N. K. Sato, N. Aso, K. Miyake, R. Shiina, P. Thalmeier, 
G. Varelogiannis, C. Geibel, F. Steglich, P. Fulde, 
and T. Komatsubara, 
Nature {\bf 410}, 340 (2001). 

\bibitem{Nakamura}
S. Nakamura, T. Moriya, and K. Ueda, 
J. Phys. Soc. Jpn. {\bf 65}, 4026 (1996).

\bibitem{Monthoux1}
P. Monthoux and G. G. Lonzarich, 
Phys. Rev. B {\bf 63}, 054529 (2001).

\bibitem{Arita}
R. Arita, K. Kuroki, and H. Aoki, 
Phys. Rev. B {\bf 60}, 14585 (1999).

\bibitem{Bickers}
N. E. Bickers, D. J. Scalapino and S. R. White, 
Phys. Rev. Lett. {\bf 62}, 961 (1989).

\bibitem{Pao}
C.-H. Pao and N. E. Bickers, 
Phys. Rev. Lett. {\bf 72}, 1870 (1994).

\bibitem{Monthoux2}
P. Monthoux and D. J. Scalapino, 
Phys. Rev. Lett. {\bf 72}, 1874 (1994).

\bibitem{Dahm1}
T. Dahm and L. Tewordt, 
Phys. Rev. B{\bf 52}, 1297 (1995).

\bibitem{footnote}
This condition for the magnetic transition is different from 
the one used in a previous paper \cite{Bickers} 
where the susceptibility is not self-consistently renormalized. 
The latter, giving a finite $T_{\rm N}$ for a 2D-systems, 
seems to overestimate the magnetic transition temperature. 
The present use of eq. (15) ensures self-consistency. 

\bibitem{Morr}
D. K. Morr and D. Pines, 
Phys. Rev. Lett. {\bf 81}, 1086 (1998).

\bibitem{Dahm2}
T. Dahm, D. Manske, and L. Tewordt, 
Phys. Rev. B{\bf 58}, 12454 (1998).

\bibitem{Takimoto1}
T. Takimoto and T. Moriya, 
J. Phys. Soc. Jpn. {\bf 67}, 3570 (1998).

\bibitem{Rossat-Mignod}
J. Rossat-Mignod, L. P. Regnault, C. Vettier, P. Bourges, 
P. Burlet, J. Bossy, J. Y. Henry, and G. Lapertot, 
Physica C {\bf 185-189}, 86 (1991).

\bibitem{Fong1}
H. F. Fong, P. Bourges, Y. Sidis, L. P. Regnault, 
A. Ivanov, G. D. Gu, N. Koshizuka, and B. Keimer, 
Nature {\bf 398}, 588 (1999).

\bibitem{Fong2}
H. F. Fong, P. Bourges, Y. Sidis, L. P. Regnault, 
J. Bossy, A. Ivanov, D. L. Milius, I. A. Aksay, 
and B. Keimer, 
Phys. Rev. B{\bf 61}, 14773 (2000).

\bibitem{Takimoto2}
T. Takimoto, 
Phys. Rev. B{\bf 62}, 14641 (2000).

\bibitem{Takimoto3}
T. Takimoto, T. Hotta, T. Maehira, and K. Ueda, 
preprint.

\end{references}
\end{document}